\documentclass[twocolumn,showpacs,preprintnumbers,amsmath,amssymb,pre]{revtex4}

\usepackage{graphicx}
\usepackage{dcolumn}
\usepackage{overpic}
\usepackage{float}

\begin{document}

\title{Topological Bogoliubov excitations in inversion-symmetric systems of interacting bosons} 

\author{G. Engelhardt}
\email{georg@itp.tu-berlin.de}

\author{T. Brandes}

\affiliation{%
Institut f\"ur Theoretische Physik, Technische Universit\"at Berlin, Hardenbergstra\ss e 36, 10623 Berlin, Germany}%

\begin{abstract}

On top of the mean-field analysis of a Bose-Einstein condensate, one typically applies the Bogoliubov 
theory to analyze quantum fluctuations of the excited modes.
Therefore, one has to diagonalize the Bogoliubov Hamiltonian in a symplectic manner.
In our article we investigate the topology of these Bogoliubov excitations in inversion-invariant systems of interacting bosons.
We  analyze how the condensate influences the topology of the Bogoliubov excitations.
Analogously to the fermionic case, here we  establish a symplectic extension of the polarization characterizing the topology of the
Bogoliubov excitations and link it to the eigenvalues of the inversion operator at the inversion-invariant momenta. 
We also demonstrate  an instructive but experimentally feasible example that this quantity is also related to edge states in the excitation spectrum.
\end{abstract}

\pacs{ 67.85.--d, 03.75.--b, 73.20.At}

\maketitle

\newcommand{\up}{\uparrow}
\newcommand{\down}{\downarrow}
\newcommand{\rAr}{\rightarrow}

\section{Introduction}
\label{sec:Introduction}
Since the discovery of Bloch bands with nontrivial topological structure, the field of topological insulators and superconductors has been  rapidly
growing~\cite{Hasan2010,Bernevig2013}.
The most prominent example is
the integer quantum Hall effect, where one can link the Hall conductance of the ground state with the Chern number of the occupied bands~\cite{Thouless1982}. 
This strict quantization  is due to the fermionic character of the particles, forcing all states within a band to be equally occupied.
For this reason, a system consisting of bosons does not exhibit such  quantized observables.

In addition, there are topologically protected edge states  as a consequence of the bulk-boundary 
relation~\cite{Hasan2010,Bernevig2013}. Thus, for noninteracting particles the  band structure determines the existence of edge states. Although noninteracting bosons condense 
in the lowest-energy mode, the wave function of the excited modes within a band can exhibit a
topological structure~\cite{Stanescu2009,Wong2013}. Therefore, there can be edge modes in the excitation 
spectrum of noninteracting bosons.

In  recent years, there  has been a great effort in  creating nontrivial topological structures of fractional quantum Hall states in 
strongly interacting bosonic systems~\cite{Yao2012,Zhu2013,Deng2014,Li2015,Grusdt2013,Kjall2012,Hormozi2012,Powell2011}. In contrast, the investigation  of the topology of  bosons
in the weakly interacting regime has received 
insignificant attention.  An interesting approach is 
discussed in Ref.~\cite{Barnett2013}, where the edge states of the Su-Schrieffer-Heeger model become dynamically unstable by properly preparing the condensate
in an excited transverse mode.
In such a setup one considers the excitations above the Bose-Einstein condensate (BEC), which are effectively single-particle-like due to an expansion 
of the Hamiltonian in orders of the condensate density~\cite{Kawaguchi2012}. The resulting Bogoliubov  Hamiltonian couples  particle excitations and 
hole excitations and has 
to be diagonalized in a symplectic manner due to the bosonic commutation relation. Therefore, the definition of a topological invariant for these Bogoliubov-Bloch
bands is \textit{a priori} not clear. 
As a result, the condensed part of the atoms has a substantial influence on the topology of the Bogoliubov excitations, which has not been discussed in the previous
literature.
There are only few  articles about the definition of a Chern number for bosonic Bogoliubov bands, though in the context of
magnonic systems \cite{Shindou2013,Shindou2013a}.

In contrast, here we  focus on the treatment of the topology in inversion-invariant systems of weakly interacting bosons in one dimension.
Our findings are relevant for  condensed matter simulations with cold atoms in optical
lattices~\cite{mazza2012optical,mazza2010emerging,goldman2013realizing,Aidelsburger2013,Miyake2013}.

In fermionic systems, the topological invariant of inversion-invariant systems is given by the macroscopic polarization~\cite{Hughes2011,Resta1994},
which is a geometric phase of the occupied bands~\cite{Zak1989}. Although its strict quantization is due to the fact that all orbitals of the bands
below the Fermi energy are equally occupied, one can also consider it to be a quantity describing the structure of the Bloch bands independent of 
the fermionic character of the particles. For this reason, for bosonic systems it can be considered to be a topological invariant not directly 
connected to a bulk observable, 
but  predicting the existence of edge states.

Here we treat the bosonic condensate within a mean-field approach. In this context we note that
Refs.~\cite{dauphin2012rydberg,raghu2008topological,weeks2010interaction,araujo2013change} study the topology of fermionic systems
using mean-field approximations.

The article is organized as follows.
In Sec.~\ref{sec:ModelSystem}, we introduce a model that is considered as an instructive but experimentally feasible  example throughout the article. 
This system has the property that
the condensate influences the topology of its Bogoliubov excitations. 
In Sec.~\ref{sec:BdGTheory} we recall Bogoliubov theory. In Sec.~\ref{sec:TopologyBdGexcitations} 
we investigate the topology of the Bogoliubov excitations.
A main result of our article is Eq.~\eqref{eq:symplecticPolarization}, which defines an extension of the macroscopic polarization for bosonic
Bogoliubov excitations.
We also discuss the problems for the definition of the polarization caused by the Goldstone mode appearing in the lowest band.
In Sec. \ref{sec:Application} we apply 
our findings to the instructive model and show the existence of edge states.

\section{Model system}

\begin{figure}[t]
  \centering
  \includegraphics[width=1.\linewidth]{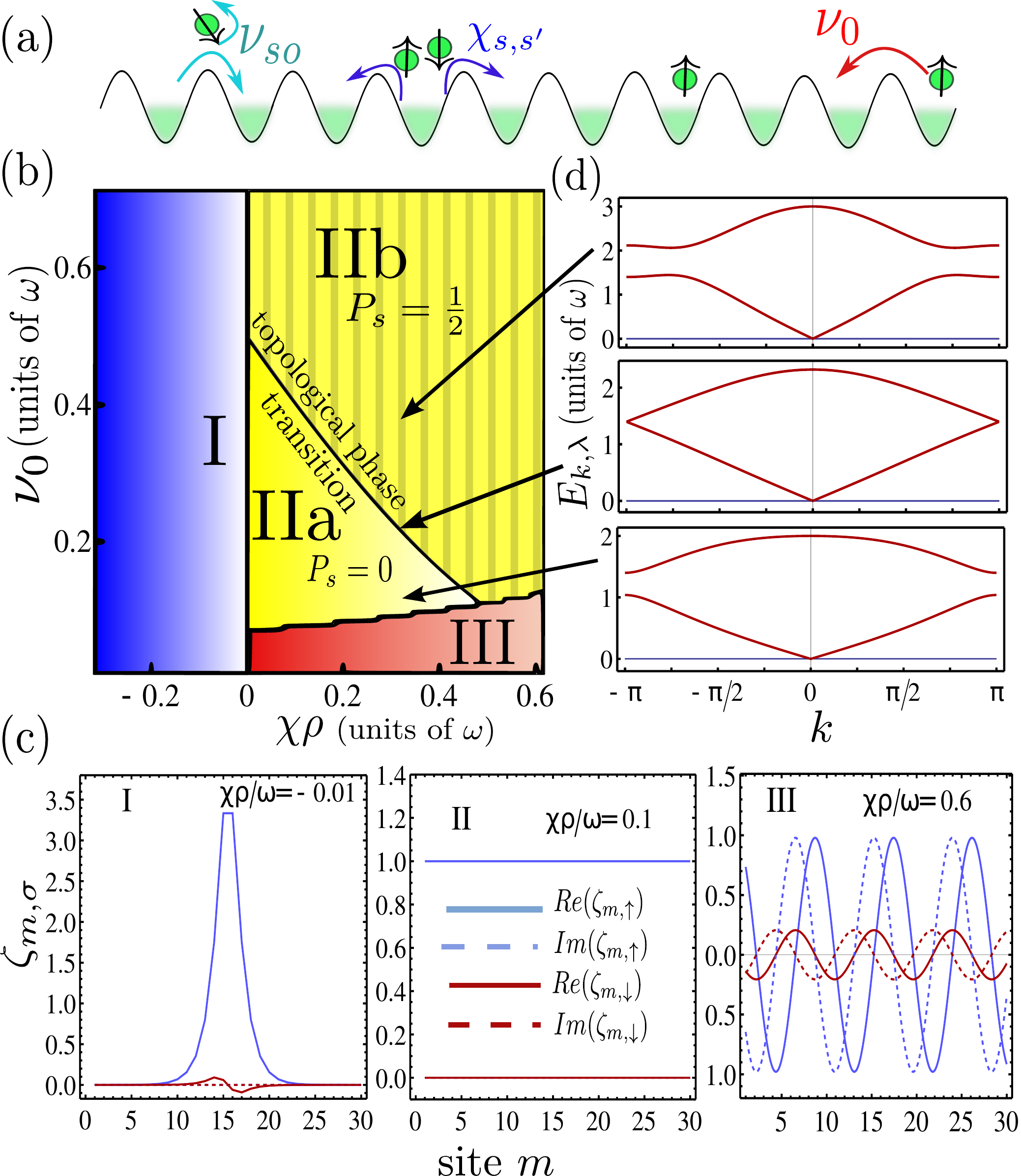}
  \caption{(Color online) (a) Sketch of the Hamiltonian~\eqref{eq:AppHamiltonian}. (b) Corresponding phase diagram distinguishing between different 
  wave functions of the condensate for the parameters $\nu_{\text{so}}/\omega=0.2$ and $\chi_{s,s'}= \delta_{s,s'}\chi$. 
  (c) Instances  of the  corresponding wave functions  for  $\nu_0/\omega=0.1$. The labels I, II, and III denote phases
  with a localized wave function, condensation at zero momentum, and condensation at finite momentum $k>0$, respectively. (d) For phase II, we depict some 
  Bogoliubov dispersion relations. The corresponding parameters are marked with arrows in the phase diagram. 
  As explained in the main text, one can distinguish the topology of the Bogoliubov excitations depending on the
  symplectic polarization $P_s$. For this reason, one can split phase II into a trivial phase IIa and a topological phase IIb. The phase
  boundary is strongly  influenced by the condensate due to  the interactions between the particles. In Sec.~\ref{sec:edgeStates} we show that topologically
  protected edge states appear  in the topological phase IIb.
  }
  \label{fig:phaseDiagram}
\end{figure}

\label{sec:ModelSystem}

\subsection{Hamiltonian}

We consider systems of weakly interacting bosons in a periodic potential.  An instance of such a system  is sketched in 
Fig.~\ref{fig:phaseDiagram}(a).
There, we consider an ensemble of bosonic atoms with internal degree of freedom (spin) confined in an array of  wells created by an optical lattice. 
In position space the Hamiltonian reads 
 \begin{equation}
 H = \sum_{m=-M/2+1}^{M/2} H_m  +V_m,
 \label{eq:AppHamiltonian}
\end{equation}
 where
\begin{align}
 H_{m} &= \omega \left( \hat a_{\down,m}^\dagger \hat a_{\down,m} - \hat a_{\up,m}^\dagger \hat a_{\up,m}  \right)  \nonumber  \\
       & - \nu_{0} \left(  \hat a_{\up,m}^\dagger \hat a_{\up,m+1}   - \hat a_{\down,m}^\dagger \hat a_{\down,m+1}  + \text{H.c.} \right)  \nonumber\\
         &-\nu_{\text{so}} \left( \hat a_{\up,m}^\dagger \hat a_{\down,m-1} -   \hat a_{\up,m}^\dagger \hat a_{\down,m+1}  +   \text{ H.c.} \right) ,\\
 V_m   &= \sum_{s,s'=\up,\down} \chi_{s,s'} \hat a_{s,m}^\dagger \hat  a_{s',m}^\dagger \hat a_{s,m}  \hat a_{s',m} .
\end{align}
Here $m$ denotes the position of the wells and $\up,\down$  denote the internal degree of freedom of the bosons. The two 
states have a level splitting of $\omega$.
The atoms can
jump between the wells, described by the parameters $\nu_0$ and $\nu_{\text{so}}$. 
The modulus of the hopping integral $\nu_0$ is equal  for the two spin components but differs in sign. The term proportional to $\nu_{\text{so}}$ denotes a 
spin-orbit coupling, which can be generated within current technology~\cite{GalitskiVictor2013,mazza2012optical}. 
Additionally, we have a state-dependent interaction that is local in position space. Therefore, we ensure  that this is not in
conflict with the inversion symmetry of the Hamiltonian.

The single-particle Hamiltonian corresponds to the systems in \cite{LiXiaopeng2013,Pan2014} discussing  fermionic  systems of cold atoms that have a nontrivial topology.
These articles  suggest possible experimental implementations for the single-particle Hamiltonian. These could be also applied to bosonic systems. Additionally, one
can control the interactions between the particles using Feshbach resonances.

\subsection{Mean-field expansion} 
 
We shift the operators

\begin{equation}
 \hat a_{\up/\down,m} \rightarrow  \sqrt{\frac{N_0}{M}} \zeta_{\up/\down,m} + \hat a_{\up/\down,m} ,
\end{equation}
where $\zeta_{\up/\down,m}\in \mathbb C $ and $N_0$ denotes the number of particles in the condensate which is assumed to
be macroscopically occupied. 
The bosonic operators now
account for quantum fluctuations on top of the condensate. We expand the Hamiltonian as
\begin{equation}
 H = \rho_0 E_{\text{GP}}  + \sqrt{\rho_0}  H^{(L)} + H^{(B)}    +  O (\rho_0^{-1/2}),
 \label{eq:mfExpansion}
\end{equation} 
where $\rho_0= N_0/{M} $ is the  density of the condensed particles.
The Hamiltonian $ H^{(L)}$ ($H^{(B)}$) depends on $\{\zeta_{\up/\down,m}\}$ and contains terms that are linear (quadratic) in  bosonic operators. 

Here $E_{\text{GP}}$ is a function of $\{\zeta_{\up/\down,m}\}$ and denotes
the Gross-Pitaevskii  functional. 
It exactly reads as \eqref{eq:AppHamiltonian} with the operators replaced by $\zeta_{\up/\down,m}$ and $\chi_{s,s'}\rightarrow \chi_{s,s'} \rho_0$.
To find the mean-field ground state, we minimize it by an appropriate choice of $\zeta_{\up/\down,m}$.
The minimization procedure is performed by using a modified ansatz  of Ref.~\cite{Li2012}. Details can be found  in Appendix~\ref{sec:MinizationGPfunc}. 

The result is depicted in the phase diagram in Fig.~\ref{fig:phaseDiagram}(b) for the 
special choice $\chi_{\up \up}= \chi_{\down \down} = \chi$ and $\chi_{\up \down}=0$.
Note that  for the mean-field treatment in Fig.~\ref{fig:phaseDiagram} we assume that nearly all bosons are  condensed, so  we  approximate $\rho_0\approx\rho=N/M$, which is the
density of all particles.

We find  three phases. In phase I appearing for $\rho_0 \chi<0$ we find that the atoms condense within a small 
area with a localized wave function~\cite{Strecker2002,Khaykovich2002}. In phase II we find a condensation at zero momentum $k=0$ and $(\zeta_{\up,m},\zeta_{\down,m})=(1,0)$. Due to the 
spin-orbit coupling and the interactions, the atoms condense at a finite momentum $k>0$ in phase III. Thus, only in phase II 
there is a mean-field wave function, which is invariant under inversion. 

At a stationary point of the Gross-Pitaevskii functional, the linear part in \eqref{eq:mfExpansion} vanishes
 and the excitations are solely governed by the quadratic Bogoliubov Hamiltonian. To respect that we work at constant
 particle number we consider $N_0$ to be an operator and  replace~\cite{Kawaguchi2012}
\begin{equation}
 \hat N_0 =  N - \sum_{m,\; s= \up,\down} \hat a_{m,s}^\dagger \hat a_{m,s}.
\end{equation}
 This leads to the appearance of an effective chemical potential $\mu_{\text{eff}}$ in $H^{(B)}$, which reads
\begin{align}
   \mu_{\text{eff}} =  \frac 1 M  \sum_m  & \omega \left(   \zeta_{\down,m}^*\zeta_{\down,m}-\zeta_{\up,m}^*\zeta_{\up,m}  \right) \nonumber \\
                        &-\nu_0 \left( \zeta_{\up,m}^*\zeta_{\up,m+1} -  \zeta_{\down,m}^*\zeta_{\down,m+1}   +\text{c.c.}\right) \nonumber \\
                        &-\nu_{\text{so}} \left( -\zeta_{\up,m}^*\zeta_{\down,m+1} +  \zeta_{\up,m}^*\zeta_{\down,m+1}  + \text{c.c.} \right) \nonumber\\
                        &+2 \rho \sum_{s,s'} \chi_{s,s'} \zeta_{s,m}^*\zeta_{s',m}^* \zeta_{s,m}\zeta_{s',m}, 
\end{align}
where $\zeta_{\up/\down,m}$ denotes now the  stationary point of the Gross-Pitaevskii function $E_{\text{GP}}$.

\subsection{Bogoliubov Hamiltonian in momentum space}

We proceed to work in phase II where $(\zeta_{\up,m},\zeta_{\down,m})=(1,0)$.  As we have there a translational-invariant
Bogoliubov Hamiltonian, we can perform a  Fourier transformation
 and obtain a Bogoliubov Hamiltonian of the form~\cite{Kawaguchi2012}
 \begin{align}
 H^{(B)}&= \frac 12 \sum_k \left(  \mathbf{\hat a}^\dagger_{k}, \mathbf{ \hat a}_{-k}   \right) 
\mathbf {H}_k
 \left(
 \begin{array}{c}
                                                                                                      \mathbf{\hat a}_{k} \\
                                                                                                     \mathbf{\hat a}_{-k}^\dagger
  \end{array}
  \right)       \nonumber \\
\mathbf {H}_k&=
\left(
 \begin{array}{cc}
  \mathbf{H}^{(0)}_k + \mathbf{H}^{(1)} & \mathbf{H}^{(2)} \\
  \left[\mathbf{H}^{(2)}\right]^*  & \left[\mathbf{H}^{(0)}_{-k} + \mathbf{H}^{(1)}    \right]^*
 \end{array}
 \right),
 \label{eq:BdGHamiltonian}
\end{align}
where $\mathbf{\hat a}_k^\dagger = \left( \hat a^\dagger_{k,\up},\hat  a^\dagger_{k, \down} \right)$ is a vector of  bosonic creation
 operators and
 \begin{align}
  \mathbf{H}^{(0)}_k&=
  \left(
  \begin{array}{cc}
                 -\omega -  2 \nu_0 \cos k      &      2 i \nu_{\text{so}} \sin k                                                      \\
             -2 i \nu_{\text{so}} \sin k                                      &       \omega +  2 \nu_0 \cos k              
  \end{array}
  \right)\label{eq:singlePartHam} \\
    \mathbf{H}^{(1)}&=
  \left(
  \begin{array}{cc}
                  4 \chi  \rho  -\mu_{\text{eff}}   &      0                                                   \\
             0                                    &            2 \chi_{\up\down} \rho -\mu_{\text{eff}}     
  \end{array}
  \right) \\
      \mathbf{H}^{(2)}&=
  \left(
  \begin{array}{cc}
             2 \chi \rho   &     0                                                \\
             0                                   &              0                     
  \end{array}
  \right).
 \end{align}
The chemical potential  reduces to $\mu_{\text{eff}}=-\omega-2 \nu_{0}+ 2 \rho \chi $. The Bogoliubov Hamiltonian determines the topological properties
of the excitations. 
In order to investigate this, one first has to diagonalize the Bogoliubov Hamiltonian. 
Importantly,
one cannot diagonalize the Bogoliubov Hamiltonian by a simple unitary 
transformation as the resulting quasiparticles would not fulfill bosonic commutation relations. 
In contrast, the diagonalization has to be performed in a symplectic manner.
Consequently, one cannot apply the  definitions of topological invariants for  usual noninteracting systems directly, but has to respect 
the symplectic nature of the diagonalization procedure. In the next
section, we briefly recall this procedure. Then we  can define a symplectic extension of 
the macroscopic polarization constituting a topological
invariant that characterizes the Bogoliubov-Bloch bands.

\section{Bogoliubov Theory}

\label{sec:BdGTheory}

A generic Bogoliubov Hamiltonian  can be written in the form 
$\mathcal{H}^{(B)} = \frac 12 \left(  \mathbf{\hat a}^\dagger ,\mathbf{\hat  a}   \right) \mathbf {H}  \left(  \mathbf{\hat a} ,\mathbf{\hat a} ^\dagger  \right)^T$,
where
\begin{align} 
\mathbf {H}=
\left(
 \begin{array}{cc}
  \mathbf{H}^{(\alpha)}  & \mathbf{H}^{(\beta)}  \\
  \left[\mathbf{H}^{(\beta)}\right]^*  & \left[ \mathbf{H}^{(\alpha)}    \right]^*
 \end{array} \label{eq:generalBdGHam}
 \right)
\end{align}
and $\mathbf{\hat a}^\dagger = \left( \hat a^\dagger_{p = 1} \cdots \hat  a^\dagger_{p = \mathcal N} \right)$.
 The label $p$ may denote the position, momentum, or
 internal degree of spinor bosons. 
The matrix $\mathbf{H}^{(\alpha)} $ is Hermitian and $\mathbf{H}^{(\beta)}$ is symmetric.
We also assume that $\mathbf H$ is positive definite.

This Hamiltonian can be diagonalized with the ansatz 
$\left(  \mathbf{\hat a}^\dagger ,\mathbf{ \hat a}   \right)= \left(  \mathbf{\hat b}^\dagger, \mathbf{\hat b}   \right) \mathbf{T}^\dagger$,
where $\mathbf{T}$ denotes a $2\mathcal N \times 2\mathcal N$ paraunitary matrix \cite{Colpa1978,Shindou2013,Kawaguchi2012}.
The new operators $\mathbf{\hat b}$ shall also fulfill bosonic commutation relations. To this end, one has to require that
\begin{equation}
 \mathbf{T}^\dagger \boldsymbol{\sigma_z} \mathbf{T} = \boldsymbol{\sigma_z}\qquad \mathbf{T}\boldsymbol{\sigma_z} \mathbf{T}^\dagger  = \boldsymbol{\sigma_z},
 \label{eq:CompletnesT}
\end{equation}
with the diagonal matrix  $\left( \boldsymbol{\sigma_z} \right)_{l,l'}= \delta_{l,l'} \sigma_l   $ and $\sigma_l=1$ for $l\leq \mathcal N$ and $\sigma_l=-1$ otherwise.
After inserting the ansatz into the Bogoliubov Hamiltonian, one easily sees, that the Hamiltonian is diagonalized if 
\begin{equation}
 \mathbf {H} \mathbf T= \boldsymbol{\sigma_z} \mathbf T
 \left(
 \begin{array}{cc}
 \mathbf E & \\
           & -\mathbf E
 \end{array}
 \right),
 \label{eq:BdGequation}
\end{equation}
where $\mathbf{E}$ denotes a diagonal matrix $\mathbf{E} = \text{diag}\left[ E_1 ,\dots, E_{\mathcal N}\right]$ with $E_i\geq 0$.  As a consequence of Eq. \eqref{eq:BdGequation}
and of the symmetric structure of $\mathbf H$, 
the paraunitary matrix $\mathbf T$ can be written in the form
\begin{equation}
 \mathbf{T}=
 \left(
 \begin{array}{cc}
  \mathbf{U} &  \mathbf{V}^*   \\
  \mathbf{V}  &  \mathbf{U}^*
 \end{array}
\right)
\label{eq:Trepresentation}
\end{equation}
with $\mathbf{U}$ and $\mathbf{V}$ being $\mathcal N \times \mathcal N$ matrices. 
In the following we denote by $\mathbf{U}$ $(\mathbf{V})$  the particle (hole) part of the excitations.
Consequently, only the first $\mathcal N$ columns of $\mathbf T$ contain
independent solutions. The other $\mathcal N$ columns resemble   exactly the same Hamiltonian as the first one, also having a positive energy. Thus,
there are only positive excitation energies.

Finally we remark that the Hamiltonian in momentum space  $\mathbf H_k$ in  Eq. \eqref{eq:BdGHamiltonian} does not necessarily 
have the form \eqref{eq:generalBdGHam}. This problem can be solved by formally combining the entries of $\mathbf H_k$ and $\mathbf H_{-k}$ in
an enlarged matrix.

\section{Topology of Bogoliubov excitations}
\label{sec:TopologyBdGexcitations}

\subsection{Symmetry considerations}

 Let us assume that there are symmetries $\mathbf  S_j$ transforming a noninteracting Hamiltonian as 
$\mathbf S_j \mathbf{H}^{(0)}_k \mathbf S_j^{-1}= \alpha \mathbf{H}^{(0)}_{\beta k} $ with $\alpha,\beta=\pm1$.
Depending on the value of
$\alpha,\beta$, the properties of $\mathbf S_j$ and the dimension of the system one finds different topological  
classes \cite{Altland1997,Chiu2013}. 
For example, for $\alpha=-\beta=1$ and $\mathbf S_j = \mathcal P = \mathcal P^{-1}= \mathcal P^\dagger$, the single-particle
Hamiltonian is
invariant under inversion, which is the focus of our article. For the noninteracting Hamiltonian \eqref{eq:singlePartHam} 
the inversion operator can be written as $\mathcal P = \sigma_z$, where $\sigma_z$ denotes the usual $2\times2$ Pauli matrix.

Next we turn our attention to the symmetry relations of $\mathbf {H}_k$. 
Thereby, our approach is to consider the symmetry operator $\mathbf S^{B}_j\equiv \boldsymbol{1}_2\otimes \mathbf S_j $, where $\mathbf S_j$ denotes
a symmetry of $\mathbf {H}_k^{(0)}$ and $\boldsymbol{1}_2$ is the $2\times2$ identity matrix.

It is natural 
that the symmetry of the Bogoliubov Hamiltonian has a block structure as otherwise the symmetry would
relate a pure particlelike excitation with a particle-hole-like one. 
Due to the appearance of the additional  matrices $\mathbf {H}_k^{(1,2)}$, the 
symmetries  $\mathbf S_j$ of $\mathbf {H}_k^{(0)}$ do  not necessarily create symmetries of $\mathbf {H}_k$,
so  this symmetry can be lost. 
Consequently, the interaction of the particles can change the topological classification.

However, let us assume that the Bogoliubov Hamiltonian and  the symplectic extension of the inversion
 symmetry $\mathcal P^{B} =  \boldsymbol{1}_2 \otimes \mathcal P$ fulfill
\begin{equation}
 \mathcal P^{B}  \mathbf{H}_k \mathcal P^{{B} } = \mathbf{H}_{-k}. 
 \label{eq:defInversionSymmetry}
\end{equation}
This condition is fulfilled for the Hamiltonian \eqref{eq:BdGHamiltonian}.
For most $k$ values, this does not impose a constraint as the symmetry just connects the Hamiltonian at $k$ with that at $-k$. 
Due to the periodicity in position space, the momentum is only defined within the Brillouin zone which we assume to have length $2\pi$. Therefore, 
the momentum $k_{\text{inv}}=0$ and the boundary of the Brillouin zone $k_{\text{inv}}=\pi$ are invariant under
inversion as $- k_{\text{inv}} \mod 2\pi= k_{\text{inv}}$.
For these momenta  relation \eqref{eq:defInversionSymmetry} exhibit a strict constraint  for the Hamiltonian $\mathbf{H}_{k_{\text{inv}}}$.

\subsection{Symplectic Polarization}

For a single-particle Hamiltonian the so-called macroscopic polarization constitutes a topological invariant~\cite{Hughes2011,Resta1994}.
We now want  to formulate a symplectic generalization of the macroscopic polarization determining the topology.

For the following derivations we consider  systems with a discrete basis. The extension to continuous systems
 works analogously, but one has to be careful with the dimension of the basis.

Let $\mathbf{T}$ be the solution of Eq.~\eqref{eq:BdGequation} in position space with periodic boundary conditions. 
The rows of $\mathbf{T}$ can be labeled with $(m,l)$, where $m\in \{-M/ 2+1,\dots,M/2\}$ denotes the position of the unit cells 
 and $l$ an internal degree of freedom within the unit cell.
The  columns of $\mathbf{T}$ contain the eigenstates of~\eqref{eq:BdGequation}.
As we consider periodic systems, they can be labeled with the indices $(k,\lambda)$, where $k=2\pi\tilde m  /M $ with $\tilde m\in \{-M/ 2+1,\dots,M/2\}$ denotes 
a quasimomentum within the Brillouin zone and $\lambda \in \{1,\dots,2 \mathcal L\}$ denotes the band index.
Let us further denote the $(k,\lambda)$th column of $\mathbf{U}$ ($\mathbf{V}$) by $T^u_{(k,\lambda)}$ 
($T^v_{(k,\lambda)}$) as being $(\mathcal N= M \times  \mathcal L)$-dimensional vectors.

Their entries can 
be expressed in terms of 
Bloch functions $\left(T^c_{(k,\lambda)}\right)_{(m,l)} =\frac1 {\sqrt{M}} e^{i k m} t^c_{(k,\lambda),l}$, where $c\in \{u,v \}$. 

The periodic parts
are the solutions of the Bogoliubov equation in momentum space 
\begin{equation}
 \mathbf {H}_k \mathbf t_{k}= \boldsymbol{\sigma_z} \mathbf t_{k}
  \left(
 \begin{array}{cc}
 \mathbf E_k & \\
           & -\mathbf E_{-k}
 \end{array}
 \right),\label{eq:BdGequationMomentum}
 \end{equation}
 where $\mathbf {H}_k$ is the matrix in \eqref{eq:BdGHamiltonian}.
 The paraunitary matrix $\mathbf t_{k}$ has dimension $2 \mathcal L \times 2 \mathcal L$. It also fulfills $\mathbf t_{k}^\dagger\boldsymbol{\sigma_z}\mathbf t_{k}=\boldsymbol{\sigma_z}$ 
 and $\mathbf t_{k}\boldsymbol{\sigma_z}\mathbf t_{k}^\dagger=\boldsymbol{\sigma_z}$. 
 More precisely, for $\lambda\leq\mathcal L$ the 
 relation reads  $(\mathbf t_k)_{l,\lambda}=t^c_{ (k,\lambda),l \mod \mathcal L } $,  with  $c=u$ for $l\leq \mathcal L$ and  $c=v$ otherwise. 
 For $\lambda>\mathcal L$ we have  $(\mathbf t_k)_{l,\lambda}=(t^{c}_{ (-k,\lambda-\mathcal L),l \mod \mathcal L})^* $,  with  $c=v$ for $l\leq \mathcal L$ and  $c=u$ otherwise~\cite{Kawaguchi2012}.
 
Analogously to the noninteracting case
we define the corresponding Wannier functions for $M \rightarrow \infty$ as 
\begin{align}
 w^c_{\lambda,m,l}= \frac 1{2\pi} \int_{\text{BZ}} dk e^{i k m } t^c_{(k,\lambda),l}\quad ,
\end{align}
where BZ denotes the Brilloin zone.
Here we explicitly distinguish between particle $c=u$ and hole $c=v$ contributions to the Wannier function.
Before defining the polarization, we have to sort the bands $\lambda$. As can be seen in Eq.~\eqref{eq:BdGequation},
we always have  pairs of energies $\pm E$. We consider here only the columns with $\lambda<\mathcal L$ corresponding to positive $E$.
Due to \eqref{eq:Trepresentation}, the columns corresponding to negative energies contain only copies of $\lambda<\mathcal L$.
We consider the bands up to 
an energy $E_{\text{max}}$ which shall be in a band gap.
Sorting the columns with $E>0$ by energy, we denote the band with the largest energy $E_{\lambda}<E_{\text{max}}$ with $\lambda_{\text{max}}$. 
Thus we consider the bands
$\lambda \leq \lambda_{\text{max}}$ so that there is an energy gap between $\lambda_{\text{max}}$ and $\lambda_{\text{max}}+1$.
For the Hamiltonian~\eqref{eq:BdGHamiltonian}, we depict some dispersion relations in Fig.~\ref{fig:phaseDiagram}(d). As
$\mathcal L =2$, we have two bands and the spectrum is gapped between $\lambda=1$ and $\lambda=2$.

For $c=u,v$ we separately  define the corresponding contributions to the macroscopic polarization to be
\begin{equation}
 P_c   \equiv  \lim_{M \rightarrow \infty }   \sum_{m = -M/2+1}^{M/2} \sum_{\stackrel{\lambda\leq\lambda_{\text{max}}}{l}} \; \; (w_{\lambda,m,l}^{c})^* \;m\;w^c_{\lambda,m,l} . \label{eq:defContPol} 
\end{equation}

With these definitions we can define the symplectic polarization as the difference of the particle  
and hole polarization contributions 
\begin{align}
 P_s \equiv  P_u - P_v  
            = \frac 1{2\pi} \int_{\text{BZ}} dk A(k) ,
            \label{eq:symplecticPolarization}
\end{align}
where we  introduced the Berry  potential
\begin{equation}
 A(k) = i\sum_{\lambda\leq \lambda_{\text{max}}}  \text{Tr}\left[ \mathbf \Gamma_\lambda \boldsymbol{\sigma_z} \mathbf{t}_k^\dagger \boldsymbol{\sigma_z} \left( \frac{\partial }{\partial k} \mathbf{t}_k\right)  \right].
  \label{eq:defVectorPotential}
 \end{equation}
 We define the matrix $\left(\mathbf \Gamma_\lambda \right)_{j,j'}= \delta_{j,j'}\delta_{j,\lambda}$ 
as being a $2\mathcal L \times 2 \mathcal L$ matrix.
The symplectic polarization
of the bands $\lambda\leq\lambda_{\text{max}}$   determines, whether or not there is an edge state between the bands $\lambda_{\text{max}}$ and $\lambda_{\text{max}}+1$.

 For the noninteracting case the symplectic polarization reduces to 
$P_s\rightarrow P_u$ and coincides with the  macroscopic polarization  of Ref.~\cite{Resta1994}.
Equation~\eqref{eq:defVectorPotential} agrees with the Berry potential of Ref.~\cite{Shindou2013} found in the 
context of a bosonic Chern number. Yet, in that article there
is no interpretation in terms of the symplectic polarization \eqref{eq:symplecticPolarization}.
A proof of the last step in \eqref{eq:symplecticPolarization} is given in Appendix \ref{app:DerivationDetails}. We also prove in
Appendix \ref{app:RaelBerryPot} that
the Berry potential is real valued.

\subsection{Topological invariant}

The symplectic polarization of inversion-invariant systems is strictly quantized to the values $P_s=m,\frac 12 + m $  with $m\in \mathbb Z$ as the  one for noninteracting
systems \cite{Hughes2011}. 
 The proof also
works essentially as in the noninteracting case, yet one has to respect the symplectic structure of the
eigenstates. We first define the sewing matrix connecting the state at $k$ with the one
at $-k$. If we have a solution of the Bogoliubov equation \eqref{eq:BdGequationMomentum} in momentum space $\mathbf t_k$,
then $\mathcal P^{B} \mathbf t_k$ diagonalizes $\mathbf{H}_{-k}$. Thus, one can connect the paraunitary matrices  at
$k$ and $-k$ as
\begin{align}
\mathbf B_k &= \mathbf t_{-k}^\dagger \boldsymbol{\sigma_z} \mathcal P^{B} \mathbf t_{k} \Leftrightarrow \\
\mathbf t_{-k} &= \mathcal P^{B}  \mathbf t_{k}\boldsymbol{\sigma_z} \mathbf B_k^\dagger,
\label{eq:defSewingMatrix}
\end{align}
where $\mathbf B_k$ denotes the sewing matrix. Importantly, it can only mix states being degenerate.
When there are no degeneracies, the sewing matrix reduces to a diagonal matrix with elements of unit modulus.
As we assumed that our system is gapped between $\lambda_{\text{max}}$ and $\lambda_{\text{max}}+1$, the sewing matrix $\mathbf{B}_k$
has a block-diagonal structure. We denote the block referring to the band below the gap by $\mathbf B_{<,k}$.
For the inversion-invariant momenta $k_{\text{inv}}$, its determinant is a product of the eigenvalues of $\mathcal P^{B}$ regarding the 
eigenstates, $\mathcal P^{B} t_{k_{\text{inv}},\lambda}=\eta_{\lambda}(k_{\text{inv}}) t_{k_{\text{inv}},\lambda}$, 
thus
\begin{equation}
 \text{det} \left[\mathbf B_{<,k_{\text{inv}}} \right]= \prod \limits_{\lambda\leq\lambda_{\text{max}}} \eta_{\lambda}(k_{\text{inv}}).
 \label{eq:DetB}
\end{equation}

The sewing matrix obeys the same transformation rules as $\mathbf t_k$,  
\begin{align}
 \mathbf B^\dagger_k  \boldsymbol{\sigma_z} \mathbf B_k &= \mathbf t_{k}^\dagger \mathcal P^{B} \boldsymbol{\sigma_z}\mathbf t_{-k}  \boldsymbol{\sigma_z} \mathbf t_{-k}^\dagger  \boldsymbol{\sigma_z} \mathcal P^{B}  \mathbf t_{k}\nonumber  \\
                                                    &=  \boldsymbol{\sigma_z}.
              \label{eq:CompletnesB}                                      
 \end{align}
Using the sewing matrix, we link the symplectic polarization $P_s$ to the eigenvalues of the symplectic 
inversion operator.
Analogously to the noninteracting case~\cite{Hughes2011}, we need to relate the Berry potential at $k$ and
$-k$, but respecting the symplectic structure of the eigenstates. We find
\begin{equation}
 A(-k)= -A(k) +i \partial_{ k} \ln \left[ \text{det}\left(  \mathbf B_{<,k} \right)  \right],
\label{eq:VectorPotInversion}
 \end{equation}
which we prove in Appendix \ref{app:BerryPotSymmetryRelation}. 
Using this we finally find
\begin{align}
 P_s &= \frac 1{2\pi}  \int_{0}^\pi dk \left[ A(k) + A(-k) \right] \\
     &=  \frac i{2\pi  }  \int_{0}^\pi dk \partial_{ k} \ln \left[ \text{det}\left( \mathbf B_{<,k} \right) \right]\\
     &= \frac i{2\pi  } \left \lbrace \ln\left[ \text{det}\left(  \mathbf B_{<,\pi}  \right) \right]- \ln\left[ \text{det}\left( \mathbf B_{<,0}  \right) \right]  \right\rbrace  \label{eq:polDerDet} \\
     &= \frac i{2\pi  } \ln \left[ \prod \limits_{\lambda \leq \lambda_{\text{max}}} \eta_\lambda(0)  \eta_\lambda(\pi) \right].
     \label{eq:AnalyticsPs}
\end{align}
In the last step we have used that the eigenvalues of the inversion operator are  $\eta_{\lambda}(k_{\text{inv}})= \pm1$. Representing the eigenvalues  in 
the form $\eta_{\lambda}(k_{\text{inv}})=1=e^{i 2\pi m }$ or $\eta_{\lambda}(k_{\text{inv}})=-1=e^{i(\pi+2\pi m)}$ finally proves that $P_s = m,\frac 12 +m$.

\subsection{Polarization of the lowest  band}

The Bogoliubov excitations of a BEC typically exhibit a Goldstone mode in the lowest band denoted here by $\lambda=1$. This means a linear dispersion for
small  $\left|k\right|$ and thus $E_{k,1} \propto \left|k\right|$, which can be seen in   Fig.~\ref{fig:phaseDiagram}(d).
The solution at $k=0$  resembles the mean-field solution $ \Psi_0$ obtained by the Gross-Pitaevskii  equation in the form 
$t_{0,1}= \left( u_{0,1},v_{0,1}\right)^T= \left( \Psi_0,-\Psi_0\right)^T$. Yet, this solution
is not normalizable according to \eqref{eq:CompletnesT} as $t_{0,1}^\dagger \boldsymbol{\sigma_z} t_{0,1}=0$. 

The fact that $t_{0,1}$ is not normalizable  is an obstruction for defining the Berry potential
in Eq.~\eqref{eq:defVectorPotential} at $k=0$. Nevertheless, we  argue here how to circumvent this obstacle. We use a slightly modified definition
for the symplectic polarization to respect the difficulties of the lowest band, namely,
\begin{equation}
  P_{s}= \frac 1{2\pi} \lim_{\delta \downarrow 0 }\left[ \int_{-\pi}^{-\delta} dk A(k) + \int_{\delta}^{\pi} dk A(k) \right].
\end{equation}
Of course, relation~\eqref{eq:VectorPotInversion} for $k\neq0$ is not affected and therefore one can adopt the  derivation up to line~\eqref{eq:polDerDet} by
including a limit operation so that
\begin{equation}
  P_{s}= \frac i {2\pi  } \left \lbrace \ln\left[ \text{det}\left( \mathbf B_\pi  \right) \right]- \lim_{\delta \downarrow 0} \ln\left[ \text{det}\left( \mathbf B_\delta  \right) \right]  \right\rbrace  .\label{eq:polDerDetGold} 
\end{equation}
The crucial point is to discuss the limit. 
We also assume that the lowest band is non-degenerate in a finite region around $k=0$. Consequently, the reduced sewing matrix has the form
\begin{equation}
 \mathbf B_{<,k}= 
 \left(
 \begin{array}{cc}
   b_{k,1}&\\
                           &\tilde {\mathbf B}_{<,k}
 \end{array}
 \right).
\end{equation}
The submatrix $\tilde {\mathbf B}_{<,k}$ behaves regularly for $k\rightarrow 0$ and does not cause any problems. 
So we just have  to discuss the implications of $b_{k,1}$.
To this end, for a given Gross-Pitaevskii solution $ \Psi_0$, we
define a normalization function $f_k >0$ so that
\begin{equation}
 \lim_{k \rightarrow  0} f_k t_{k,1} = \left( \Psi_0, -\Psi_0\right)^T.
\end{equation}
The exact shape of $f_k$ is not crucial for our discussion, but in agreement with inversion symmetry we require $f_k = f_{-k}$. 
 For $k \rightarrow 0$,
the relation between the solutions at momenta connected by inversion reads, according to \eqref{eq:defSewingMatrix},
\begin{equation}
  t_{-k,1}= e^{i b_{k,1}}  \mathcal P^{B}  t_{k,1}.
\end{equation}

This relation is not well defined at $k=0$. Therefore, we multiply $f_k$ so that the limit $k \rightarrow 0$ exists on both sides of the 
equation. Consequently,
\begin{equation}
 e^{i b_{0,1}} \mathcal P^{B} = \mathbf 1
\end{equation}
 and this constrains $b_{0,1}= 2 \pi  m$ or   $b_{0,1}=  \pi (2 m+ 1)$ depending on the eigenvalue of $ \mathcal P^{B}$  regarding
 $\left(  \Psi_0,-\Psi_0\right)^T$. Conclusively one can say  that, although the  limit $k\rightarrow 0$ of the state
 is not well defined, the phase is up to $2\pi m$, so   the final outcome in \eqref{eq:AnalyticsPs} is not affected.

 \subsection{Application}
 
 \label{sec:Application}
 \begin{figure}[t]
  \centering
  \includegraphics[width=0.49\linewidth]{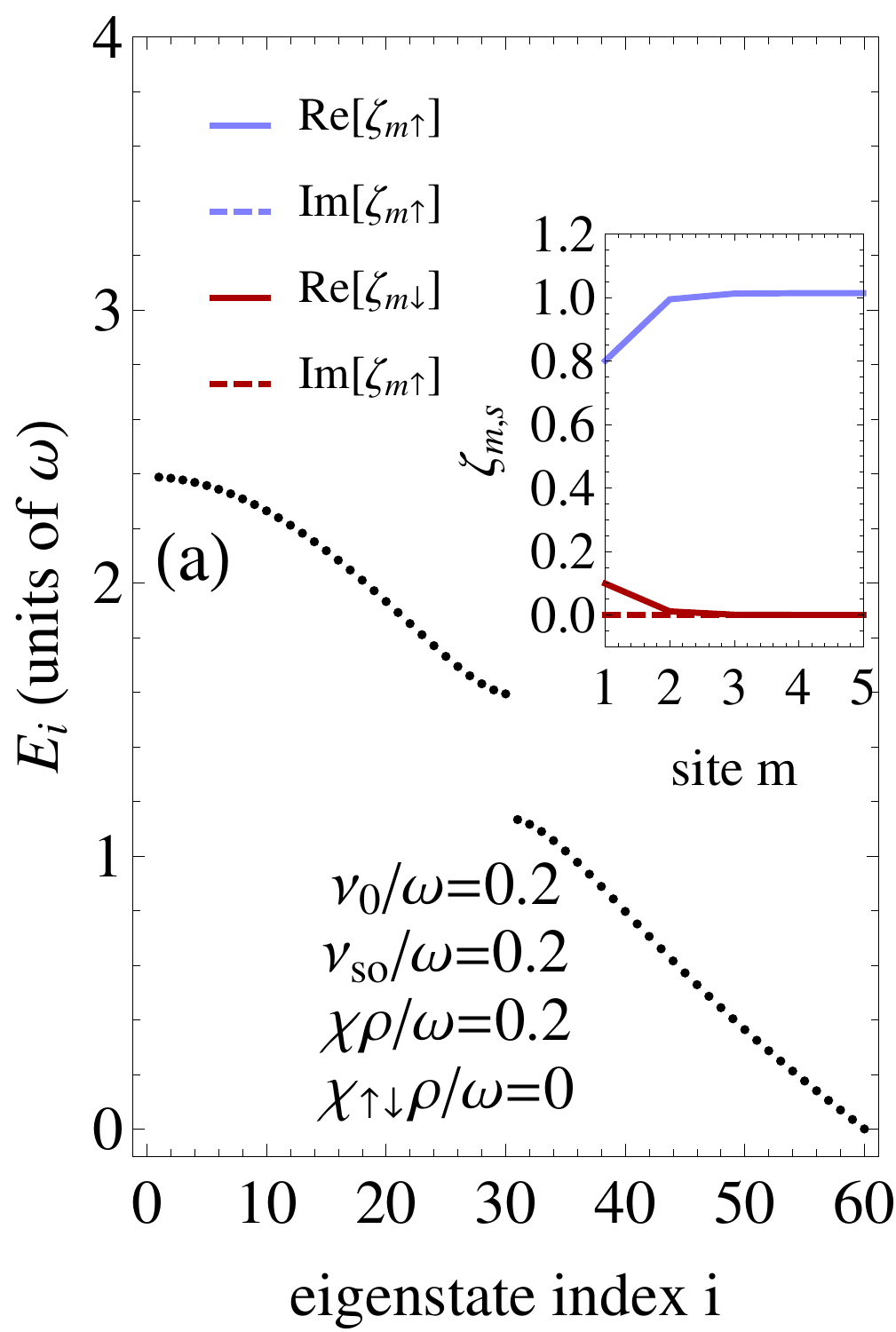}
  \includegraphics[width=0.49\linewidth]{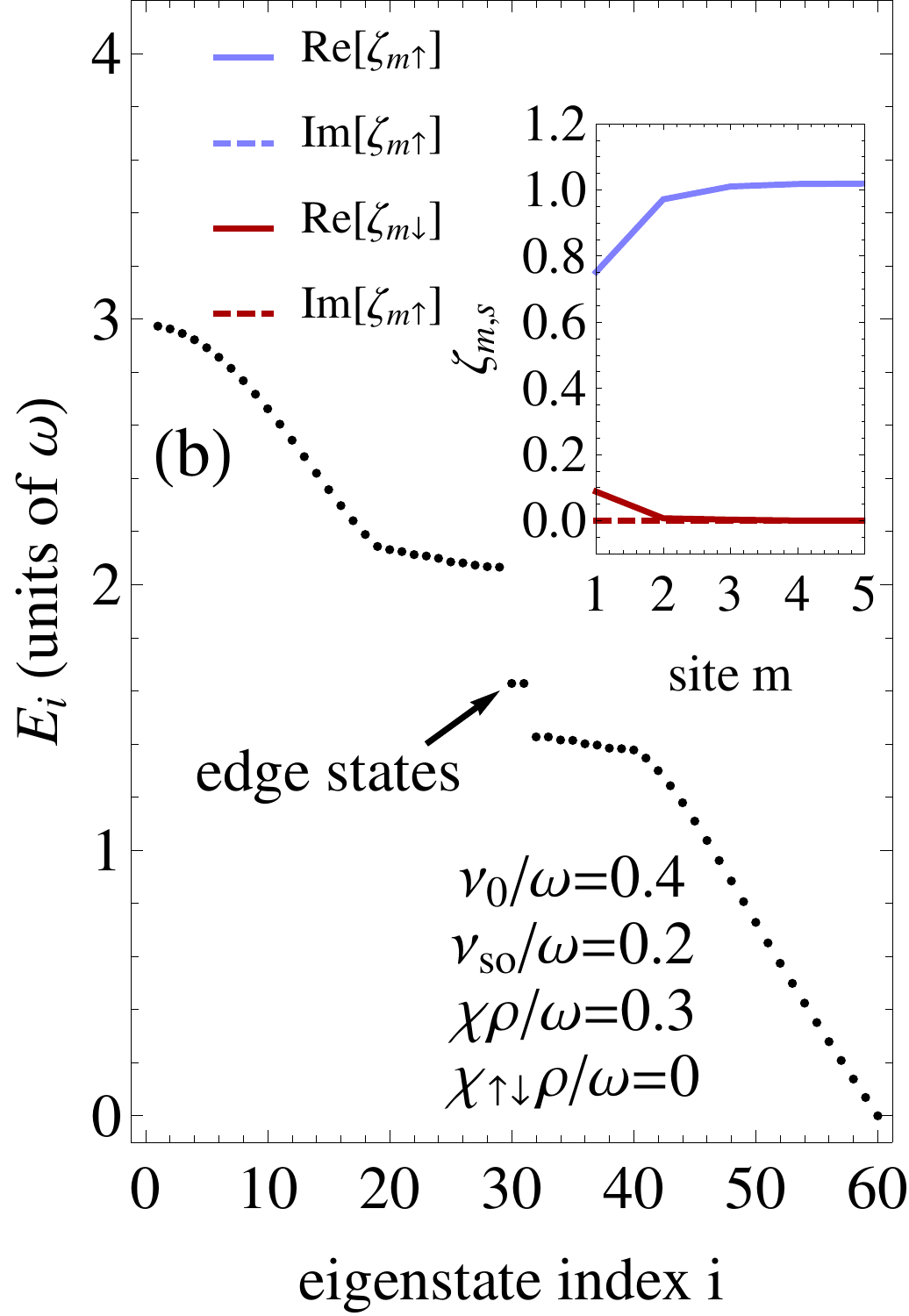}
   \includegraphics[width=1.\linewidth]{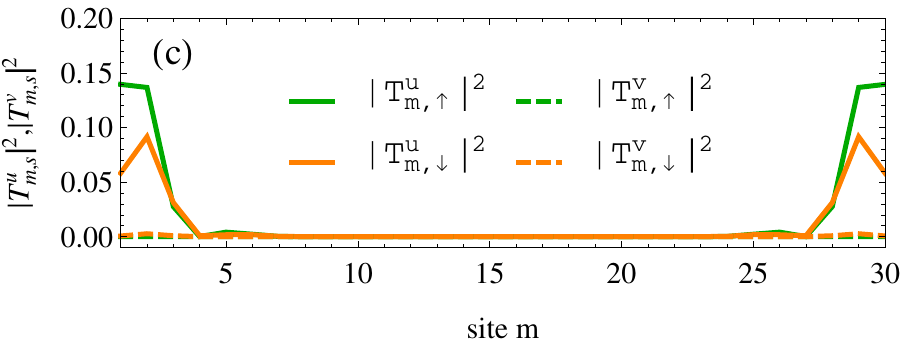}
  \caption{(Color online) (a) and (b) Spectra of the Bogoliubov excitations with fixed boundary conditions for $M=30$ sites in phases IIa and IIb,
  respectively. Before performing the Bogoliubov diagonalization,
  we first minimize the Gross-Pitaevskii functional. The resulting wave function of the condensates close to the boundaries is depicted in the insets. Due to the fixed boundary condition, the density of the condensate
  is lower at  the boundaries. Using the condensate wave function, we then perform the Bogoliubov diagonalization. The excitation energies are 
  all positive and real valued, as we have prepared the wave function in its ground state. One clearly identifies the midgap states. (c) Wave function
  of one of these midgap states.
  }
  \label{fig:edgeStates}
\end{figure}
 
 Here we return to the Hamiltonian \eqref{eq:BdGHamiltonian} and investigate its topology.
As  the Bogoliubov Hamiltonian and the matrix $\boldsymbol{\tau}_z= \mathbf 1_2 \otimes \sigma_z$ fulfill
\begin{equation}
 \boldsymbol{\tau}_z \mathbf {H}_k \boldsymbol{\tau}_z =  \mathbf {H}_{-k},
\end{equation}
 the operator $\boldsymbol{\tau}_z$ is an inversion symmetry. In Fig.~\ref{fig:phaseDiagram}(d) we depict some dispersion relations of 
the system.

To determine the topology of our model we have to consider the inversion-invariant momenta $k_{\text{inv}}=0,\pi$.
Let us consider the lower band.
As the mean-field wave function $(\zeta_{m,\up},\zeta_{m,\down})=(1,0)$ is constant for the parameters in phase II,
the eigenvalue under inversion is $\eta_{k=0,1}=1$.
The gap closes just at the boundary of the Brillouin zone so that the symplectic polarization $P_s$  can only change there.
To investigate this, we consider the eigenstates at $k=\pi$, which read 
\begin{align}
 E_{\pi,1} &= 4 \sqrt{\nu_0(\rho \chi +\nu_0)},  \quad E_{\pi,2} = 2(\omega-\rho \chi+\rho \chi_{\up\down}) ,\\
t_{\pi,1} &\propto 
\left(
 \begin{array}{c}
  \rho \chi+ 2\nu_0 - 2 \sqrt{\nu_0(\rho \chi +\nu_0)}\\
  0 \\
  \rho \chi \\
  0
 \end{array}
 \right),
  t_{\pi,2} = 
 \left(
 \begin{array}{c}
  0\\
  1 \\
  0 \\
  0
 \end{array}
 \right).
\end{align}
Obviously,  they fulfill $\boldsymbol{\tau}_z t_{\pi,\lambda}= -(-1)^{\lambda}t_{\pi,\lambda}$. Consequently,
the topological invariant~\eqref{eq:AnalyticsPs} changes at the degeneracy point $E_{\pi,1}= E_{\pi,2}$. Thus, the boundary
between the topological phases is 
\begin{multline}
  \nu_{0,\text{tpt}} =  \\
     \frac {  -\rho \chi + \sqrt{2 (\rho \chi)^2 -2 \rho \chi (\omega+\rho\chi_{\up\down})+ (\omega+\rho\chi_{\up\down})^2 } }{2}  .
\end{multline}
We depict this topological phase boundary also in the phase diagram in Fig.~\ref{fig:phaseDiagram}(b) for $\chi_{\up\down}=0$.
 One can see, that the product $\rho\chi$
has a strong impact on the topology of the system. The topological invariant for the lower band is $P_s=0$ for $\nu_0< \nu_{0,\text{tpt}}$ and
changes to $P_s=\frac 12$ for $\nu_0> \nu_{0,\text{tpt}}$. Accordingly, the system is in a topologically trivial phase or a  nontrivial phase, respectively.

\subsection{Edge states}
 
 \label{sec:edgeStates}
 Although we have defined a topological invariant, there is still the question about the physical consequences of it.
  In contrast to fermionic systems, where the polarization is an actual physical quantity, in  bosonic systems this is
  not the case as not all momenta of a band are equally occupied. However, the symplectic polarization~\eqref{eq:symplecticPolarization} determines
  the existence of edge modes in a system with finite length and fixed-boundary conditions.

  As a demonstration, we consider our instructive model  in Fig.~\ref{fig:edgeStates}. For this illustration, we 
  do not assume an additional harmonic confining potential. Due to its topological origin, the edge states are robust in the presence of
   moderate perturbations~\cite{Buchhold2012}. We also refer to Ref.~\cite{Goldman2010} for the creation of sharp boundaries.
  
  Importantly, first we have to determine the mean-field wave function by 
  minimizing the Gross-Pitaevskii functional. Here the Gross-Pitaevskii mean field has no uniform density at all sites due to the boundaries. We depict 
  the part of the condensate close to the left boundary in the insets of Figs.~\ref{fig:edgeStates}(a) and (b). 
  One can see that the density is smaller close to the edges.
  Away from the boundaries,
  the mean field is approximately constant, so  the results derived in Sec.~\ref{sec:TopologyBdGexcitations} are still valid. 
  
  On top of the mean-field wave function, we perform a Bogoliubov diagonalization in position space and depict its spectrum in
  Figs.~\ref{fig:edgeStates}(a) and (b).
  In  Fig.~\ref{fig:edgeStates}(a) the system is in the trivial phase IIa, so  no midgap states appear.
  In contrast, one clearly  identifies two midgap states within the spectrum in Fig.~\ref{fig:edgeStates}(b) depicting the spectrum for parameters in phase IIb.
  In Fig.~\ref{fig:edgeStates}(c) we plot 
  the wave function of one of these
  states. We find that it is strongly localized on the edges. The interpretation of these edge states works analogously to that in 
  the fermionic case~\cite{Hughes2011}. There, each edge state contributes  half an electron to each boundary. Thus, one particle
  splits into  two half particles. In the bosonic case, correspondingly, each edge mode can be considered to  consist of two half modes at the boundaries.
  
\section{Conclusions }

We investigated the topology of Bogoliubov excitations in inversion-invariant systems of interacting bosons.
To characterize the topology, we extended the definition of the macroscopic polarization in a symplectic manner.
We called this  quantity symplectic polarization which is defined in Eq.~\eqref{eq:symplecticPolarization} as the difference of the particlelike and
holelike polarization contributions. As in  noninteracting systems with inversion symmetry, this quantity can be expressed  by the inversion
eigenvalues of the states at inversion-invariant momenta. In an instructive example we showed, that the topological 
invariant strongly depends on the condensate density, so  the interaction between the particles modifies the 
topology of the excitations. 

The definition of the symplectic polarization can also be applied to analyze inversion-invariant systems  in higher 
dimensions. In this case we expect that an invariant defined as the product of the inversion eigenvalues of 
the states at the inversion-invariant momenta predicts edge states \cite{Hughes2011}. Furthermore,  one can link also
the symplectic polarization of the one-dimensional system to the Chern number in two dimensions~\cite{Shindou2013}.
Another possible application is in inversion-invariant systems consisting of arrays of  (pseudo)spins with large 
angular momentum such as in  Refs.~\cite{Shindou2013a,Sorokin2014,Zou2014}, where fluctuations above the mean-field treatment
are diagonalized by Bogoliubov theory.

A drawback of the bosonic edge modes in the excitation spectrum is that they are weakly occupied. However, this obstacle could be
circumvented by engineering a bosonic system in a driven setup within Floquet theory analogously to~\cite{Benito2014}.

Importantly, the  symplectic polarization discussed here can be used to 
define the symplectic generalization of the time-reversal polarization \cite{Fu2006}. This can be used to analyze the topology of time-reversal 
invariant systems of interacting bosons.

\begin{acknowledgments}
The authors gratefully acknowledge financial support
from the DFG (Germany) Grants No. BR 1528/7, No. BR 1528/8,
No. BR 1528/9, No. SFB 910, and No. GRK 1558, and inspiring conversations with V. M. Bastidas.
\end{acknowledgments}

 \bibliography{topology}

\appendix

\section{Minimization of the Gross-Pitaevskii functional}

\label{sec:MinizationGPfunc}

The Gross-Pitaevskii functional reads
 \begin{equation}
E_{\text{GP}} = \sum_{m=-M/2+1}^{M/2} E_{0,m}  +E_{V,m},
 \label{eq:GrossPitaevski}
\end{equation}
 where
\begin{align}
 E_{0,m} &= \omega \left( \zeta_{\down,m}^* \zeta_{\down,m} - \zeta_{\up,m}^* \zeta_{\up,m}  \right)   \nonumber   \\
 				&- \nu_{0} \left(   \zeta_{\up,m}^* \zeta_{\up,m+1} - \zeta_{\down,m}^* \zeta_{\down,m+1}  + \text{c.c.} \right) \nonumber \\
        			 &-\nu_{\text{so}} \left( \zeta_{\up,m}^* \zeta_{\down,m-1} -   \zeta_{\up,m}^* \zeta_{\down,m+1}  +   \text{ c.c.} \right) ,\\
 E_{V,m}  &= \sum_{s,s'=\up,\down} \rho_0 \chi_{s,s'} \zeta_{s,m}^*  \zeta_{s',m}^* \zeta_{s,m}  \zeta_{s',m} 
 .
\end{align}
Motivated by Ref.~\cite{Li2012}, we use  the modified ansatz
\begin{align}
 \left( \begin{array}{c}
          \zeta_{\up,m} \\
          \zeta_{\down,m}
          \end{array}
          \right)
			      =  C_1 \left( \begin{array}{c}
								    \cos \theta\\
								    -i \sin \theta
								  \end{array}  \right) e^{i k  m } + C_2 \left( \begin{array}{c}
														      \cos \theta\\
														      i \sin \theta
														    \end{array}  \right) e^{- i k  m } .
\end{align}
The variational parameters are $C_1$, $C_2$, $\theta$, and $k$. As we work at a fixed particle number, we have to respect the constraint 
$\left|C_1\right|^2+ \left|C_2\right|^2=1$.
Due to this ansatz the noninteracting part of the energy functional reads
\begin{align}
 &\sum_m E_{0,m} = \nonumber \\
                  &=  M \cos 2 \theta  \left(-w-  2 \nu_0 \cos k \right) + M 2 \nu_{\text{so}} \sin 2 \theta \sin k .
\end{align}
Importantly, the noninteracting part does not depend on the coefficients $C_1$ and $C_2$ which reflects the inversion symmetry of the system.
Accordingly, the interaction terms turn out to be
\begin{align}
  \sum_m E_{V,m} 
             &= M  \rho_0 \left( 1+ 2 \beta \right) \left(\chi_{\up \up} \cos^4\theta +  \chi_{\down\down} \sin^4\theta \right) \nonumber \\
               & + M \rho_0 2 \chi_{\up\down} \left( 1- 2 \beta \right)  \cos^2\theta \sin^2\theta ,
\end{align}
where  we define $\beta = \left|C_1\right|^2\left|C_2\right|^2$ with $\beta \in (0,\frac 14)$. We immediately see that $E_{\text{GP}}$ is  a 
linear function of $\beta$. Therefore, the minimum can only be located at $\beta=0$ or  $\beta=1/4$.

Next we derive a relation between $k$ and $\theta$. To this end  we take the derivative of  $E_{\text{GP}}$ with respect to $k$.
After a short algebraic manipulation, we obtain
\begin{equation}
 \tan 2 \theta = - \frac{\nu_0}{\nu_{\text{so}}} \tan k  .
\end{equation}
Having done these analytical preparations, we  can numerically minimize the Gross-Pitaevskii functional, which is now just
a function of essentially one variable, thus $E_{\text{GP}}= E_{\text{GP}}(\theta,\beta)$ as $\beta \in \{0,\frac 14\}$. The result is 
depicted in Fig.~\ref{fig:phaseDiagram}(b). For comparison, we also directly minimize the Gross-Pitaevskii functional numerically, where
we also found the localized ground-state wave function for $\chi \rho_0 <0$.

\section{Berry  potential}

\subsection{Details of the derivation}
\label{app:DerivationDetails}

We start with the final expression of the Berry potential and perform the 
proof from the end. First, we insert the representation \eqref{eq:Trepresentation} for $\mathbf{t_k}$ and 
perform the multiplications with $\boldsymbol{\sigma_z}$ so that we obtain
\begin{align}
 &A(k) = i\sum_{\lambda\leq \lambda_{\text{max}}}\text{Tr}\left[ \mathbf \Gamma_\lambda \boldsymbol{\sigma_z} \mathbf{t}_k^\dagger \boldsymbol{\sigma_z} \left( \partial_k \mathbf{t}_k \right)  \right] \\
          &= i\sum_{\lambda \leq \lambda_{\text{max}}}\text{Tr} \left[ \mathbf \Gamma_\lambda 
            \left(
            \begin{array}{cc}
             \mathbf{u}_k^\dagger & -\mathbf{v}_k^\dagger \\
             -\mathbf{v}_{-k}^T & \mathbf{u}_{-k}^T \\
            \end{array}
            \right)
            \frac{\partial }{\partial k} 
            \left(
            \begin{array}{cc}
             \mathbf{u}_k & \mathbf{v}_{-k}^* \\
             \mathbf{v}_{k} & \mathbf{u}_{-k}^* \\
            \end{array}
            \right)
             \right]\\
          &= i\sum_{\lambda\leq \lambda_{\text{max}}}\left( \mathbf{u}_k^\dagger \partial_k \mathbf{u}_k  -   \mathbf{v}_k^\dagger \partial_k  \mathbf{v}_k \right)_{\lambda,\lambda}     .
\end{align}
We evaluate the matrix product by inserting a complete $\mathbf 1$ of the basis states $l$ of the unit cell  so that
\begin{align}
 A(k) =  i  \sum_l \; \sum_{\stackrel{\lambda\leq \lambda_{\text{max}}}{c=u,v}} \sigma(c) \,(t_{ (k,\lambda),l}^{c})^*\:\partial_k\:t^c_{(k,\lambda),l} ,   
 \end{align}
using $\sigma(u)=+1$  and $\sigma(v)=-1$.
We have also used that the columns of $\mathbf u_k$ and $\mathbf v_k$ are the periodic parts of the Bloch function, namely, $t^u_{(k,\lambda),l}$ and
$t^v_{(k,\lambda),l}$, respectively. We continue by inserting   unity  so that
\begin{widetext}
\begin{align}
\frac 1 {2\pi} \int_{\text{BZ}} dk A(k) &=   \frac {i}{2\pi} \int_{\text{BZ}} dk  dk' \sum_l \; \sum_{\stackrel{\lambda\leq \lambda_{\text{max}}}{c=u,v}}  \sigma(c) \ (t^{c}_{(k',\lambda),l})^*\: \delta(k-k') \partial_k\:  t^c_{(k,\lambda),l} \\
                                 &=\frac {i}{(2\pi)^2} \lim_{M \rightarrow \infty}\int_{\text{BZ}} dk  dk' \sum_l \; \sum_{\stackrel{\lambda\leq \lambda_{\text{max}}}{c=u,v}}  \sum_{m=-M/2+1}^{M/2} \sigma(c) \, (t^{c}_{(k',\lambda),l})^*\: e^{i (k-k')m}  \partial_k\:t^c_{(k,\lambda),l} \label{eq:latticePoints}\\
                                 &=-\frac {i}{(2\pi)^2} \lim_{M \rightarrow \infty}\int_{\text{BZ}} dk  dk' \sum_{m,l} \; \sum_{\stackrel{\lambda\leq \lambda_{\text{max}}}{c=u,v}}  \sigma(c) \, (t^{c}_{(k',\lambda),l})^*\:  \left(\partial_ke^{i(k-k')m} \right)\:t^c_{(k,\lambda),l} \\
                                 &=-\frac {i}{(2\pi)^2} \lim_{M \rightarrow \infty}\int_{\text{BZ}} dk  dk' \sum_{m,l} \; \sum_{\stackrel{\lambda\leq \lambda_{\text{max}}}{c=u,v}}  \sigma(c) \,  (t^{c}_{(k',\lambda),l})^*\:  e^{-i k'm} \: i \:m \:t^c_{(k,\lambda),l} e^{i k m}\\
                                 &= P_u-P_v.
\end{align}
\end{widetext}
This finally proves Eq.~\eqref{eq:symplecticPolarization}.
\clearpage

\subsection{Real valueness of the Berry potential}
\label{app:RaelBerryPot}

To prove that the Berry potential \eqref{eq:defVectorPotential} is real valued,
we calculate
\begin{align}
  A^*(k) &=- i \sum_{\lambda\leq \lambda_{\text{max}}}\left \lbrace \text{Tr}\left[ \mathbf{\Gamma}_\lambda \boldsymbol{\sigma_z} \mathbf{t}_k^\dagger \boldsymbol{\sigma_z} \left( \partial_k \mathbf{t}_k\right)  \right] \right\rbrace^* \\
           &= - i  \sum_{\lambda\leq \lambda_{\text{max}}} \text{Tr}\left[ \left( \partial_k \mathbf{t}_k\right)^\dagger \boldsymbol{\sigma_z}  \mathbf{t}_k  \boldsymbol{\sigma_z}  \mathbf{\Gamma}_\lambda \right]  \\
          &= -i  \sum_{\lambda\leq \lambda_{\text{max}}}\text{Tr}\left[ \left( \partial_k \mathbf{t}_k^\dagger \right)\boldsymbol{\sigma_z}  \mathbf{t}_k  \boldsymbol{\sigma_z}  \mathbf{\Gamma}_\lambda \right]  \\
          &= i  \sum_{\lambda\leq \lambda_{\text{max}}} \text{Tr}\left[ \mathbf{t}_k^\dagger  \boldsymbol{\sigma_z} \left( \partial_k \mathbf{t}_k\right) \boldsymbol{\sigma_z}  \mathbf{\Gamma}_\lambda   \right]  \\
          &= A(k).
 \end{align}

\subsection{Symmetry relation}

\label{app:BerryPotSymmetryRelation}
Here we prove relation \eqref{eq:VectorPotInversion}. By definition we have
\begin{align}
&A(-k) =  i \sum_{\lambda\leq \lambda_{\text{max}}}\text{Tr} \left[ \mathbf  \Gamma_\lambda \boldsymbol{\sigma_z} \mathbf{t}_{-k}^\dagger \boldsymbol{\sigma_z} \left( \partial_{-k} \mathbf{t_{-k}}\right)  \right]\\
           &= i\sum_{\lambda} \text{Tr} \left[ \mathbf \Gamma_\lambda  \mathbf B_k \boldsymbol{\sigma_z}  \mathbf{t}_{k}^\dagger \mathcal P^{B}                         \boldsymbol{\sigma_z} \left( \partial_{ -k} \mathcal P^{B}  \mathbf t_{k} \boldsymbol{\sigma_z} \mathbf B_k^\dagger \right)  \right]\\
           &= i\sum_{\lambda} \text{Tr} \left[ \mathbf \Gamma_\lambda    \mathbf B_k \boldsymbol{\sigma_z}  \mathbf{t}_{k}^\dagger \mathcal P^{B}                \boldsymbol{\sigma_z}  \mathcal P^{B} \left( \partial_{ -k}  \mathbf t_{k}  \right) \boldsymbol{\sigma_z} \mathbf B_k^\dagger   \right]\\
           &+  i\sum_{\lambda} \text{Tr} \left[ \mathbf \Gamma_\lambda    \mathbf B_k \boldsymbol{\sigma_z}  \mathbf{t}_{k}^\dagger \mathcal P^{B}               \boldsymbol{\sigma_z} \mathcal P^{B}  \mathbf t_{k} \boldsymbol{\sigma_z}\left( \partial_{ -k}  \mathbf B_k^\dagger \right)   \right] \\
          &=-  A(k) -i\sum_{\lambda} \text{Tr}\left[\mathbf \Gamma_\lambda \boldsymbol{\sigma_z} \mathbf{B}_k\boldsymbol{\sigma_z} \partial_{ k} \mathbf{B}_k^\dagger  \right] .\\
           &=-  A(k)-i \text{Tr}\left[ \mathbf{B}_{<,k} \partial_{ k} \mathbf{B}_{<,k}^\dagger  \right]
           \end{align}
 
Next we  recognize that
\begin{equation}
 \boldsymbol{\sigma_z} \mathbf B_k \boldsymbol{\sigma_z} \mathbf B_k^\dagger = \mathbf 1
\end{equation}
due to Eq.~\eqref{eq:CompletnesB}. 
As the sewing matrix has a block diagonal structure, we find
\begin{equation}
  \mathbf B_{<,k}  \mathbf B_{<,k}^\dagger = \mathbf 1.
\end{equation}
This relation means that $\mathbf B_{<,k}$ is unitary. We expand it in terms of its eigenstates $\left|i \right>$ such that
\begin{equation}
 \mathbf B_{<,k}= \sum_{i} e_i \left|i \right> \left<  i \right|  \Leftrightarrow  \mathbf B_{<,k }^\dagger= \sum_{i} e_i^{-1} \left|i \right> \left<  i \right|.
\end{equation}
Both the eigenstates and the eigenvalues depend on $k$.
For a notational reason we suppress the index in the following.
Now we can prove relation \eqref{eq:VectorPotInversion}: 
\begin{align}
&\text{Tr}\left[ \mathbf{B}_{<,k} \partial_{ k} \mathbf{B}_{<,k}^\dagger  \right] \nonumber \\
&=  \text{Tr}\left[  \sum_{i} e_i \left|i \right> \left<  i \right|   \left(\partial_{ k} \sum_{j} e_j^{-1} \left| j \right> \left<  j \right|  \right) \right] \\
  &=  \text{Tr}\left[  \sum_{i} e_i \left|i \right> \left<  i \right|    \sum_{j}\left(\partial_{ k}  e_j^{-1} \right)\left|j \right> \left<  j \right|   \right] \nonumber \\ 
  &+\text{Tr}\left[  \sum_{i} e_i \left|i \right> \left<  i \right|  \sum_{j} e_j^{-1}  \left(\partial_{ k}  \left|j \right> \right)\left<  j \right|  \right] \nonumber \\
  &+\text{Tr}\left[  \sum_{i} e_i \left|i \right> \left<  i \right|   \sum_{j} e_j^{-1} \left|j \right> \left(\partial_{ k}  \left<  j \right|  \right) \right] 
\end{align}
Evaluating the traces in the eigenbasis of $\mathbf B_{<,k} $ we find 
\begin{multline}
\text{Tr}\left[ \mathbf{B}_{<,k} \partial_{ k} \mathbf{B}_{<,k}^\dagger  \right] \\
  =  \sum_{i} \partial_{ k} \ln e_i^{-1} + 
    \sum_{i} \left<  i \right|\partial_{ k} \left|i \right>   +\left( \partial_{ k}  \left<   i \right| \right) \left|i \right>             \\
  =  - \partial_{ k} \ln \prod_{i} e_i + \sum_{i} \left(\partial_{ k} \left<   i \right|\left.i \right>  \right) \\
  = - \partial_{ k} \ln \left[\det \left( \mathbf B_{<,k} \right) \right].
\end{multline}
In the derivation we have also used that $\left| e_i \right|=1$, so that $\partial_k e^{-1}_i=-\partial_k e_i$.

\end{document}